\documentclass[]{spie}  
\usepackage[]{graphicx}
\usepackage{subfigure}
\usepackage{color}
\usepackage{amsmath}

\title{Modeling and characterization of the SPIDER half-wave plate} 

\author{Sean~A.~Bryan\supit{a}*, 
Peter~A.~R.~Ade\supit{b},
Mandana~Amiri\supit{c}, Steve~Benton\supit{d},
Richard~Bihary\supit{a}, James~J.~Bock\supit{e,f},
J.~Richard~Bond\supit{g}, Joseph~A.~Bonetti\supit{f},
H.~Cynthia~Chiang\supit{h}, 
Carlo~R.~Contaldi\supit{i}, Brendan~P.~Crill\supit{e,f}, 
Daniel~O'Dea\supit{i}, Olivier~Dore\supit{e,f}, Marzieh~Farhang\supit{d}, 
Jeffrey~P.~Filippini\supit{e}, Laura~Fissel\supit{d}, Natalie~Gandilo\supit{d},
Sunil~Golwala\supit{e}, Jon~E.~Gudmundsson\supit{h},
Matthew~Hasselfield\supit{c}, Mark~Halpern\supit{c}, Kyle R. Helson\supit{a},
Gene~Hilton\supit{j}, Warren~Holmes\supit{f},
Viktor~V.~Hristov\supit{e}, Kent~D.~Irwin\supit{j},
William~C.~Jones\supit{h}, Chao~Lin~Kuo\supit{k},
Carrie~J.~MacTavish\supit{l},
Peter~Mason\supit{e}, Tracy~Morford\supit{e},
Thomas~E.~Montroy\supit{a}, C.~Barth~Netterfield\supit{d},
Alexandra~S.~Rahlin\supit{h}, Carl~D.~Reintsema\supit{j}, Daniel~Riley\supit{a},
John~E.~Ruhl\supit{a}, Marcus~C.~Runyan\supit{e},
Matthew~A.~Schenker\supit{e}, Jamil~Shariff\supit{d},
Juan~Diego~Soler\supit{d}, Amy~Trangsrud\supit{e},
Rebecca~Tucker\supit{e}, Carole~Tucker\supit{b}, and
Anthony~Turner\supit{f}
\skiplinehalf
\supit{a}Department of Physics, Case Western Reserve University,
Cleveland, OH, USA; \\
\supit{b}School of Physics and Astronomy, Cardiff University, Cardiff, UK; \\
\supit{c}Department of Physics and Astronomy, University of British
Columbia, Vancouver, BC, Canada; \\
\supit{d}Department of Physics, University of Toronto, Toronto, ON,
Canada; \\
\supit{e}Department of Physics, California Institute of Technology,
Pasadena, CA, USA; \\
\supit{f}Jet Propulsion Laboratory, Pasadena, CA, USA; \\
\supit{g}Canadian Institute for Theoretical Astrophysics, University
of Toronto, Toronto, ON, Canada; \\
\supit{h}Department of Physics, Princeton University, Princeton, NJ, USA; \\
\supit{i}Department of Physics, Imperial College, University of
London, London, UK; \\
\supit{j}National Institute of Standards and Technology, Boulder, CO, USA; \\
\supit{k}Department of Physics, Stanford University, Stanford, CA, USA; \\
\supit{l}Kavli Institute for Cosmology, University of Cambridge, Cambridge, UK }

\authorinfo{*E-mail: sean.bryan@case.edu, Tel: 1 216 368 1153}

\begin{document} 
  \maketitle 

\begin{abstract}
{\sc Spider} is a balloon-borne array of six telescopes that will observe the Cosmic Microwave Background. The 2624 antenna-coupled bolometers in the instrument will make a polarization map of the CMB with approximately one-half degree resolution at 145 GHz. Polarization modulation is achieved via a cryogenic sapphire half-wave plate (HWP) skyward of the primary optic. We have measured millimeter-wave transmission spectra of the sapphire at room and cryogenic temperatures. The spectra are consistent with our physical optics model, and the data gives excellent measurements of the indices of A-cut sapphire. We have also taken preliminary spectra of the integrated HWP, optical system, and detectors in the prototype {\sc Spider} receiver. We calculate the variation in response of the HWP between observing the CMB and foreground spectra, and estimate that it should not limit the {\sc Spider} constraints on inflation.
\end{abstract}



\section{Introduction}

Constraining inflation with high-precision measurements of Cosmic Microwave Background (CMB) polarization at degree scales is an important frontier in cosmology\cite{cmbtaskforce}. A gravity wave background generated by inflation in the early universe would produce B-mode polarization patterns in the CMB at large angular scales, which are a ``smoking gun'' of inflation. A detection of B-modes at large angular scales would be strong evidence that inflation occurred. During a long-duration balloon flight, the {\sc Spider} instrument\cite{crill08,filippini10} will map the CMB polarization at the angular scales relevant for inflation with unprecedented sensitivity.

{\sc Spider} consists of an array of 6 telescopes in a single liquid helium cryostat\cite{gudmundsson10} with a total of 2624 polarized bolometers operating at frequency bands centered at 90 GHz, 145 GHz, and 280 GHz. Periodically changing the angle of the cryogenic half-wave plate (HWP) mounted skyward of each telescope's primary optic will rotate the polarization sensitivity of the instrument. Rotating the polarization sensitivity during flight will improve the fidelity of the reconstruction of the polarized signal. Since a HWP rotates the sensitivity of the instrument without rotating the beam on the sky, it will also mitigate systematic effects due to beam asymmetry.

\section{Measuring HWP Transmission Spectra}

A HWP made from a single slab of birefringent material is designed so the optical path length difference $(n_s~-~n_f)~\times~d$ between waves polarized along the ordinary and extraordinary axes is exactly a half wavelength. This phase delay causes linearly-polarized light to rotate as it passes through the plate. An ideal HWP would rotate the polarization state of light by the same angle at any frequency. However, a real HWP made from birefringent material is only a perfect polarization rotator at a single frequency. To understand the non-idealities of a real HWP over the frequency bandpasses of the {\sc Spider} detectors, we need to verify the spectral properties of the birefringent material and its anti-reflection (AR) coating. For {\sc Spider}, the HWPs will be constructed from 330 mm diameter slabs of birefringent single-crystal sapphire, and a quarter-wave fused quartz AR coat will be applied to each side. In this paper, we describe the characterization of a prototype HWP for {\sc Spider}'s 145 GHz band, constructed of 3.05 mm thick sapphire, and .280 mm thick fused quartz AR-coatings.

\begin{figure*}
\begin{center}
\includegraphics[width=0.75\textwidth]{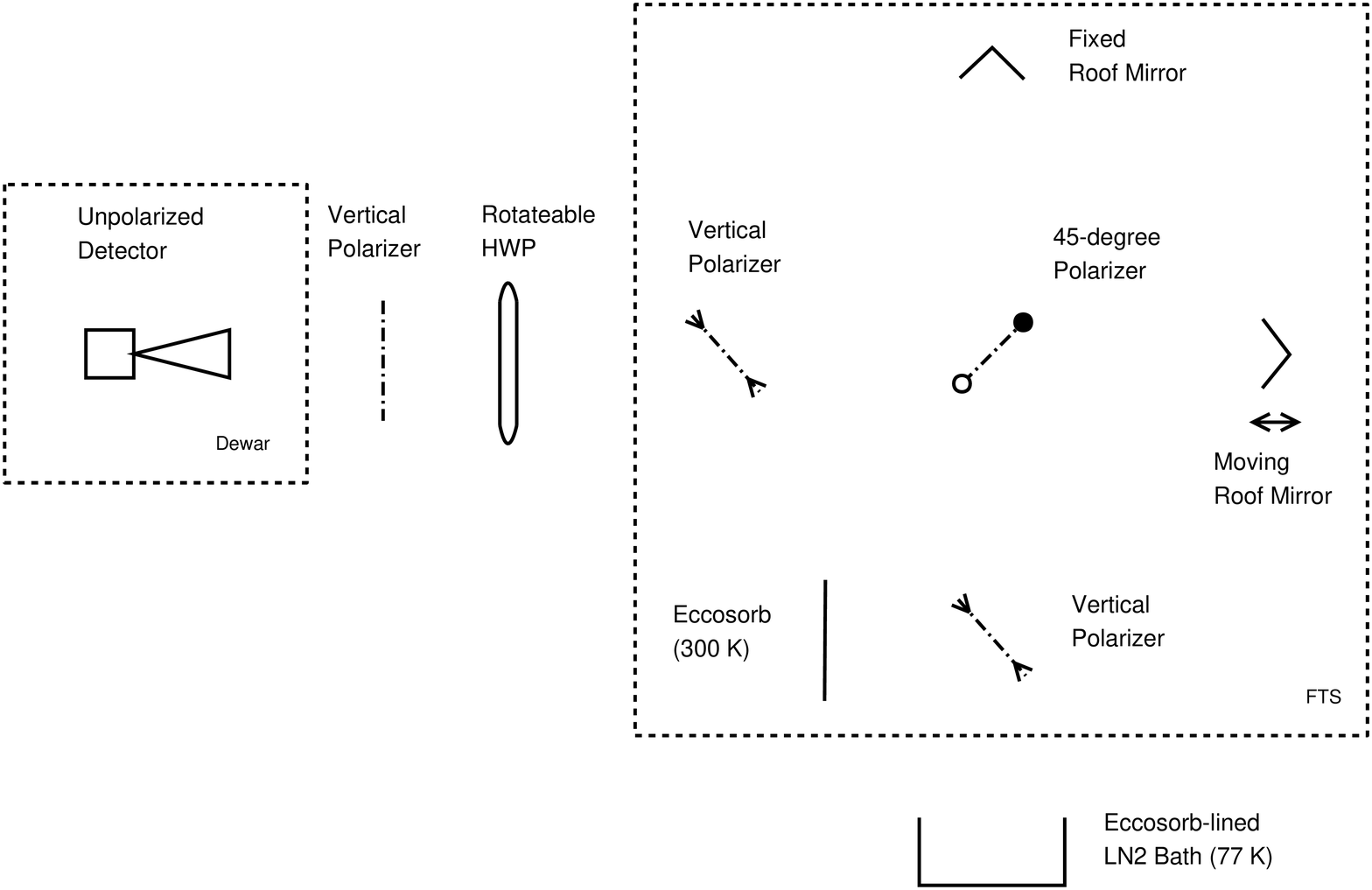}
\caption{Instrument configuration for the broadband spectra of the sapphire. We used a polarized Fourier Transform Spectrometer to shine polarized light through the sapphire slab. An analyzer polarizer at the other side of the sapphire is aligned with the output of the FTS. We used a broadband sub-kelvin bolometer as the detector to get HWP spectra at .75 GHz resolution from 100 GHz to 240 GHz. For the cryogenic spectra of sapphire, we removed the final vertical polarizer, moved the sapphire into the dewar, and used a rotatable polarizer at the output of the FTS to take spectra at several angles near both crystal axes of the sapphire. \label{fts_setup}}
\end{center}
\end{figure*}

To characterize the frequency response of the {\sc Spider} HWPs, we took broadband transmission spectra of sapphire using a polarized Fourier Transform Spectrometer (FTS). A FTS is a Michelson interferometer with a moving mirror. A frequency-independent beam splitter is difficult to make for millimeter waves, so we use a polarizing wire grid instead to construct a Martin-Pupplet interferometer \cite{lesurf90}. A block diagram of our apparatus is shown in Figure~\ref{fts_setup}.  Since the light travels down and back both arms of the interferometer, moving one mirror by a half-wavelength changes the optical path difference between the two arms by a full wavelength. This means that the mirror movement required to generate a full period in the detector timestream is $\Delta x_{mirror} = \frac{\lambda}{2}$. If the mirror is moving at a constant speed $v_{mirror}$, then $\Delta x_{mirror} = \Delta t ~v_{mirror}$. This signal will appear in the detector timestream at a frequency $f_{audio} = 1/ \Delta t$. Combining these relations yields the relationship between the audio frequency $f_{audio}$ of the signal in the detector timestream and its corresponding photon frequency $\nu$,
\begin{equation} \label{freq_conv}
\left( \frac{c}{2 v_{mirror}}\right) f_{audio} = \nu.
\end{equation}
The Fourier transform of the timestream is the product of the source, detector and HWP spectra, with the frequency axis scaled according to Equation \ref{freq_conv}. We use the discrete cosine transform for our analysis to reduce noise biasing. The frequency resolution of the FTS is determined by the maximum distance the mirror moves from the white light fringe. The data presented here was taken with a total mirror travel of .2 m, and therefore a frequency resolution of $c / (2 \times .2~\mathrm{m}) = .75$ GHz.

For the source at the input port of the FTS, we use an Eccosorb-lined liquid nitrogen bath, which emits a nearly Rayleigh-Jeans spectrum in our frequency band. We put a wire grid polarizer at the output of the FTS, followed by the sapphire HWP on a rotatable bearing, which allowed us to take a polarized spectrum at each rotation angle $\theta$ of the material. Another wire grid polarizer aligned with the output polarizer on the FTS was placed between the HWP and the detector. This allows us to see how the sapphire rotates the polarization state of the light passing through it. For the detector, we used a broadband sub-kelvin bolometer sensitive from 100 GHz to 240 GHz mounted in a lab cryostat. 

\begin{figure*}
\begin{center}
\subfigure{\includegraphics[width=0.35\textwidth]{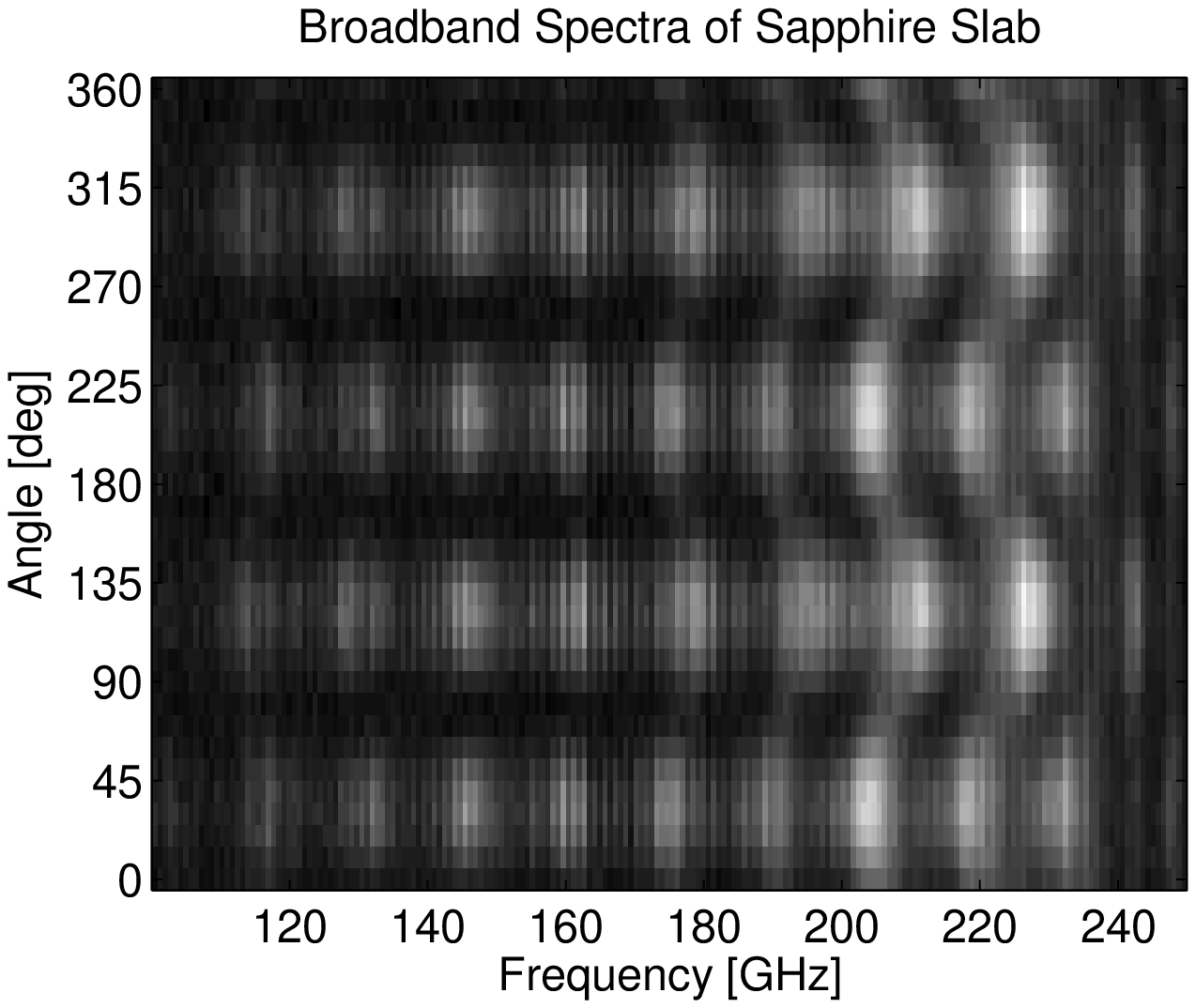}} 
\subfigure{\includegraphics[width=0.62\textwidth]{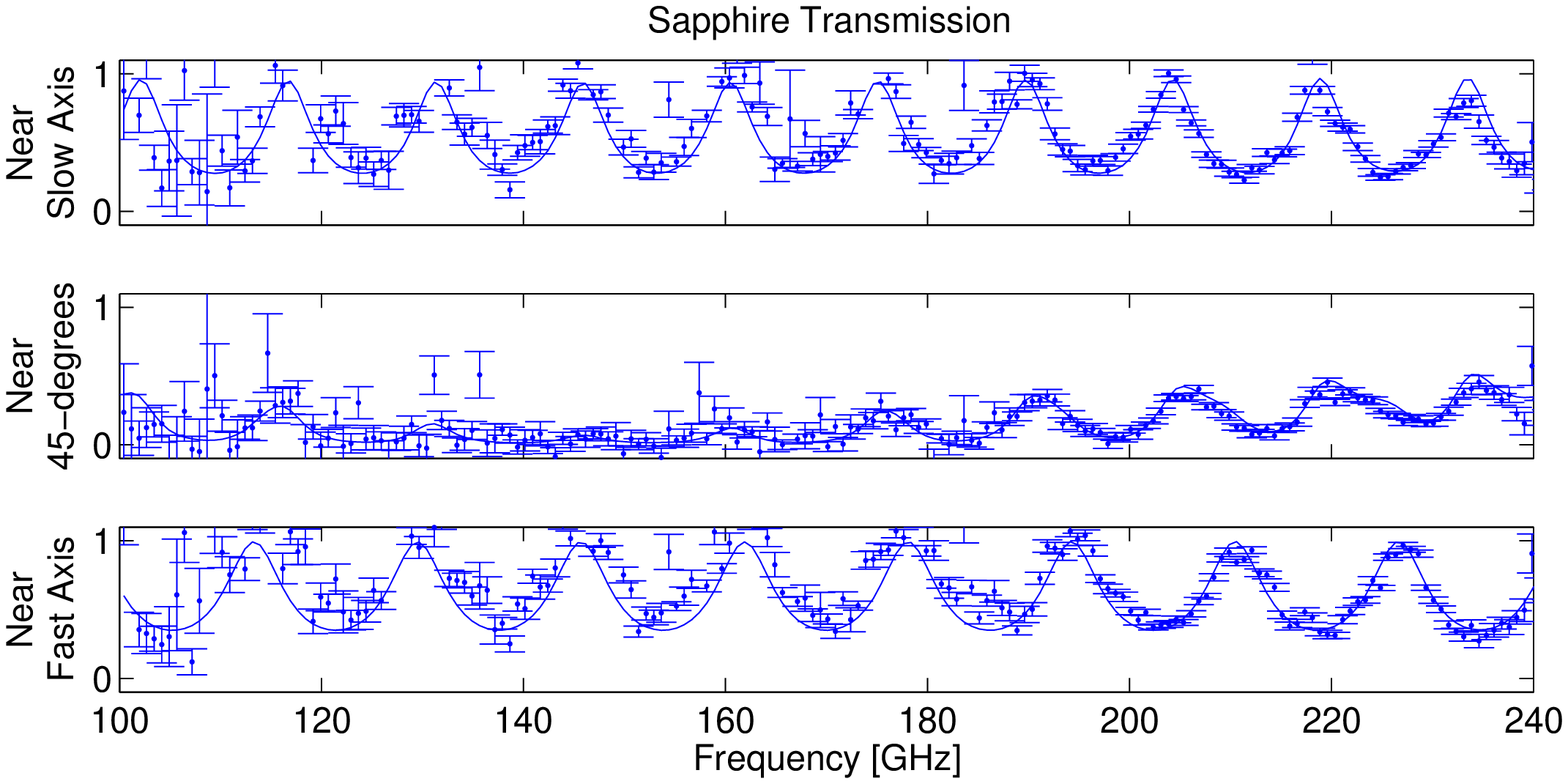}} 
\caption{Broadband millimeter-wave spectra of a 3.05 mm thick sapphire slab at room temperature. The left panel shows the observed spectra as function of angle, where the intrinsic detector response has not been removed. We took polarized spectra with the sapphire oriented at $10^{\circ}$ intervals. The variation with angle is caused by the birefringent sapphire rotating the incident linearly-polarized light from the FTS. The right panel shows spectra taken near both of the crystal axes of the birefringent sapphire, and a spectrum taken at an intermediate angle. The intrinsic detector response was removed from this plot, and the best-fit spectrum is shown as a smooth curve. The error bars on the data points are from detector noise only. The frequency spacing between the peaks in the spectrum taken at an angle near the slow axis of the sapphire is different than the peak spacing for the spectrum taken near the fast axes. This allows a precise estimate of both indices of refraction of sapphire. \label{sapphire_spectra}}
\end{center}
\end{figure*}

The transmission spectra of a 100 mm diameter sample of A-cut sapphire at room temperature are plotted in Figure~\ref{sapphire_spectra}. The slab has an air-medium interface on each side, both of which cause reflections. At certain frequencies, the interference between the two reflected waves adds constructively, causing a minimum in the transmission spectrum. At other frequencies, the interference is destructive and leads to a peak in the transmission spectrum. The frequency spacing between peaks depends on the index of refraction the polarized wave experiences as it travels through the material. For waves traveling along the slow axis of the sapphire, the spacing corresponds to the slow index of refraction of sapphire $n_s$, while for waves polarized along the fast axis, the spacing corresponds to the fast index $n_f$. We use spectra taken with the light traveling at many angles through the material, and fit them to a model. The fitting procedure implicitly uses the peak spacing and amplitude to estimate the two indices of refraction of the sapphire.

To fit our observed spectra as a function of sample angle $S_{obs}(\nu,\theta)$, we calculate the transmission $T_{xx}$ through the sapphire slab and aligned polarizers using a physical optics model similar to the one described in Savini et. al.\cite{savini06}. The model extends the 2-by-2 matrix formalism reviewed in Hecht and Zajac \cite{hecht74} for modeling multiple layers of isotropic materials to a 4-by-4 matrix formalism for multiple layers of birefringent materials. The model uses the electromagnetic boundary conditions at each of the air-material interfaces to map the incident electric and magnetic fields onto the transmitted fields. This fully treats multiple reflections and interference effects. The model can handle lossy materials, but here we assume that all materials are lossless.

Once we have a model for the HWP transmission, we multiply it by the detector response spectrum $F(\nu)$ and an overall normalization factor $a$ to obtain a model for the set of observed FTS spectra,
\begin{equation}
S_{calc}(\nu , \theta) = a \times F(\nu) \times T_{xx}(\nu, \theta - \theta_{0} \mid  \{ n_s, n_f\}),
\end{equation}
where $\theta_0$ is the angle of the crystal axes relative to the angle of the incident polarized light. We then fit the data for the parameters $\{ n_s, n_f, \theta_0, a\}$ using a Monte-Carlo Markov Chain,\cite{chib95} which allows for non-gaussian likelihood, and gives a straightforward estimate of covariances in the parameter estimates. The covariance between parameter estimates is less than 1\% for the room-temperature measurements reported here, so we ignore it below. The phase of the modulation with HWP angle determines the angle of the crystal axis, and the spacing in frequency between spectrum peaks determines the indices. When we fit spectra of an AR-coated HWP, the width of the good transmission band around the design frequency and the shape of the out of band fringes constrain the index of the AR-coating.

The room temperature sapphire spectra of our 100 mm sample and curves from the best-fit model are shown in Figure~\ref{sapphire_spectra}. The best-fit indices of refraction are shown in Table~\ref{uniformity}. Our values are almost 1\% below the values listed in Lamb\cite{lamb95}, but index differences $(n_{s}-n_{f})$ are in agreement. 
Also listed in the Table are index values derived from temperature measurements of a 330~mm diameter sapphire used in the AR-coated {\sc Spider} prototype discussed below.  For that larger sapphire, we made measurements of a roughly 50~mm diameter patch at the center, and 
three similar patches centered at a radius of $\sim 80$~mm to test uniformity of the material.  
As shown in the Table, the best fit indices for all 4 locations on the 330~mm sample agreed within errors, and agreed with the values from the 100 mm sample. 

\begin{table}
\begin{center}
\caption{Room temperature and cryogenic indices of refraction of A-cut birefringent sapphire near 145 GHz \protect\footnotemark. Statistical error from detector noise is shown next to each index value, and our estimated systematic uncertainties are shown in the last row. We measured the indices of our 330 mm diameter sample by taking spectra through its center, and at three other locations. The large sample has uniform optical properties at room temperature within statistical error. \label{uniformity}}
\begin{tabular}{ c | c | c | c | c | c |}
 & $\mathbf{n_s}$ & $\mathbf{n_f}$ & $\mathbf{n_{s} - n_{f}}$ \\
\hline
\hline
\textbf{Room Temp.} & & &\\
\hline
Lamb\cite{lamb95}&$3.403\phantom{4} \pm .003\phantom{0}$& $3.069\phantom{4} \pm .003\phantom{0}$& $.334\phantom{4} \pm .004\phantom{0}$   \\
\hline
100 mm Diameter &$3.3736 \pm .0002$& $3.0385 \pm .0002$& $.3350 \pm .0004$\\
\hline
330 mm Diameter & & &\\
\hline
\textit{Center}&$3.3742 \pm .0003$& $3.0373 \pm .0003$& $.3369 \pm .0004$   \\
\hline
\textit{Location 1}&$3.372\phantom{4} \pm .002\phantom{4}$& $3.031\phantom{4} \pm .002\phantom{4}$& $.341\phantom{4} \pm .003\phantom{4}$  \\
\hline
\textit{Location 2}&$3.371\phantom{4} \pm .002\phantom{4}$& $3.033\phantom{4} \pm .003\phantom{4}$& $.338\phantom{4} \pm .004\phantom{4}$   \\
\hline
\textit{Location 3}&$3.370\phantom{4} \pm .002\phantom{4}$& $3.030\phantom{4} \pm .003\phantom{4}$& $.340\phantom{4} \pm .004\phantom{4}$   \\
\hline
\hline
\textbf{LHe Temp.} & & &\\
\hline
Loewenstein\cite{loewenstein73} (1.5 K)&$3.361\phantom{4 \pm .0004}$& $3.047\phantom{4 \pm .0004}$& $.314\phantom{4 \pm .0004}$   \\
extrapolated\cite{johnson04} to 150 GHz & & & \\
\hline
5 K, 100 mm Diameter &$3.336\phantom{4} \pm .003\phantom{4}$& $3.019\phantom{4} \pm .003\phantom{4}$& $.317\phantom{4} \pm .004\phantom{4}$   \\
\hline
\hline
\textbf{Est. Systematics} &$\phantom{3.3734} \pm .003\phantom{4}$&$\phantom{3.3734} \pm .002\phantom{4}$& \\
\hline
\end{tabular}
\end{center}
\end{table}

 \footnotetext{Sapphire supplied by Crystal Systems, Salem, MA.}

\begin{table}
\begin{center}
\caption{Estimated systematic error budget. All of the systematics we considered cause a fractional change in the observed indices. The increase in optical path due to a tilted sample and observing through a finite aperture can only bias the observed indices higher than the true value. We re-ran our fitting code with the polarizers in the model rotated by $\pm2^{\circ}$ to determine the effect of mis-aligned polarizers on the results. The mirror motion varies at the $\pm2~\mu$m/s level due to the stage encoder resolution and motor drive feedback error. Added in quadrature, these effects are comparable to detector noise for our cold measurements, and dominate over detector noise for our warm measurements. \label{systematics}}
\begin{tabular}{ c | c | c|}
\textbf{Effect} & \textbf{Index Error} & Comment \\
\hline
\hline
Tilted Sample &  $+\phantom{.00}0\%$ &$\pm 2.5^{\circ}$ tilt.\\
 & $-.008\%$ &  \\
 \hline
Expanding Beam & $\pm.03\% $&Uncertainty in  correction \\
& & for f/3.3 optics. \\
\hline
Polarizer alignment & $\pm .01\%$& $\pm 2^{\circ}$ rotation\\
\hline
Mirror Speed & $\pm .07\%$& $\pm2~\mu$m/s for 3 mm/s motion.\\
\hline
\hline
\textit{Quadrature Sum} & $\pm.08\%$& Detector noise is $\pm .006\%$ at 300 K,\\
& & \phantom{00000......}and $\pm .09\%$ at 5 K. \\
\end{tabular}
\end{center}
\end{table}

Previous measurements of the indices of sapphire at liquid helium temperatures, such as those of Loewenstein et. al.\cite{loewenstein73} from .9 THz to 9 THz at 1.5 K, demonstrate that cryogenic indices are shifted from their room temperature values. 
Johnson\cite{johnson04} extrapolated these shifts from the high frequencies of Loewenstein et. al. down to 150 GHz. 
To check this extrapolation, we cooled our 100 mm diameter sapphire sample to 5~K, and fit the spectra to obtain its indices of refraction. Rather than rotating the sample at 5~K, we left it in a fixed position and took spectra with the FTS polarized at angles very near the slow and fast indices of the material. Our measured cold indices, listed in Table~1, differ by nearly 1\% from the extrapolated values calculated by Johnson, but our index difference is in agreement. 


To estimate the systematic uncertainties of our method, we considered four effects:
\begin{itemize}
\item Tilt in the sample mount.  The optical path of a ray traveling through a tilted HWP is larger than the optical path at normal incidence;  any tilt of the HWP biases our observed indices above their true values.
\item The non-parallel beam.  The HWP is positioned in the converging f/3.3 beam between the FTS and the detector.  Similar to the tilt effect listed above, the average non-normal incidence affects the measured indices.  We correct for this effect and assign a conservative 50\% uncertainty in the correction due to the uncertain illumination profile.
\item Polarizer misalignment.  To estimate the effect of a misalignment of the two polarizers bracketing the HWP, we ran our fitting code with mis-aligned polarizers in the physical optics model to see how the derived indices changed.
\item FTS mirror speed.  The FTS position linear encoder indicates the velocity varies at the $\pm2~\mu$m/s level for the 3~mm/sec speed used while taking the data here.  Since the peak spacing affects the indices, this couples directly to an uncertainty in the derived indices.
\end{itemize}
The systematic uncertainties associated with these effects, along with their quadrature sum, are listed in Table~\ref{systematics}.  Mirror speed is the dominant effect, and leads to a systematic uncertainty that is greater than the statistical uncertainties for our room temperature measurements, and comparable to the statistical uncertainty for the 5~K measurement.

\begin{figure*}
\begin{center}
\subfigure{\includegraphics[width=0.40\textwidth]{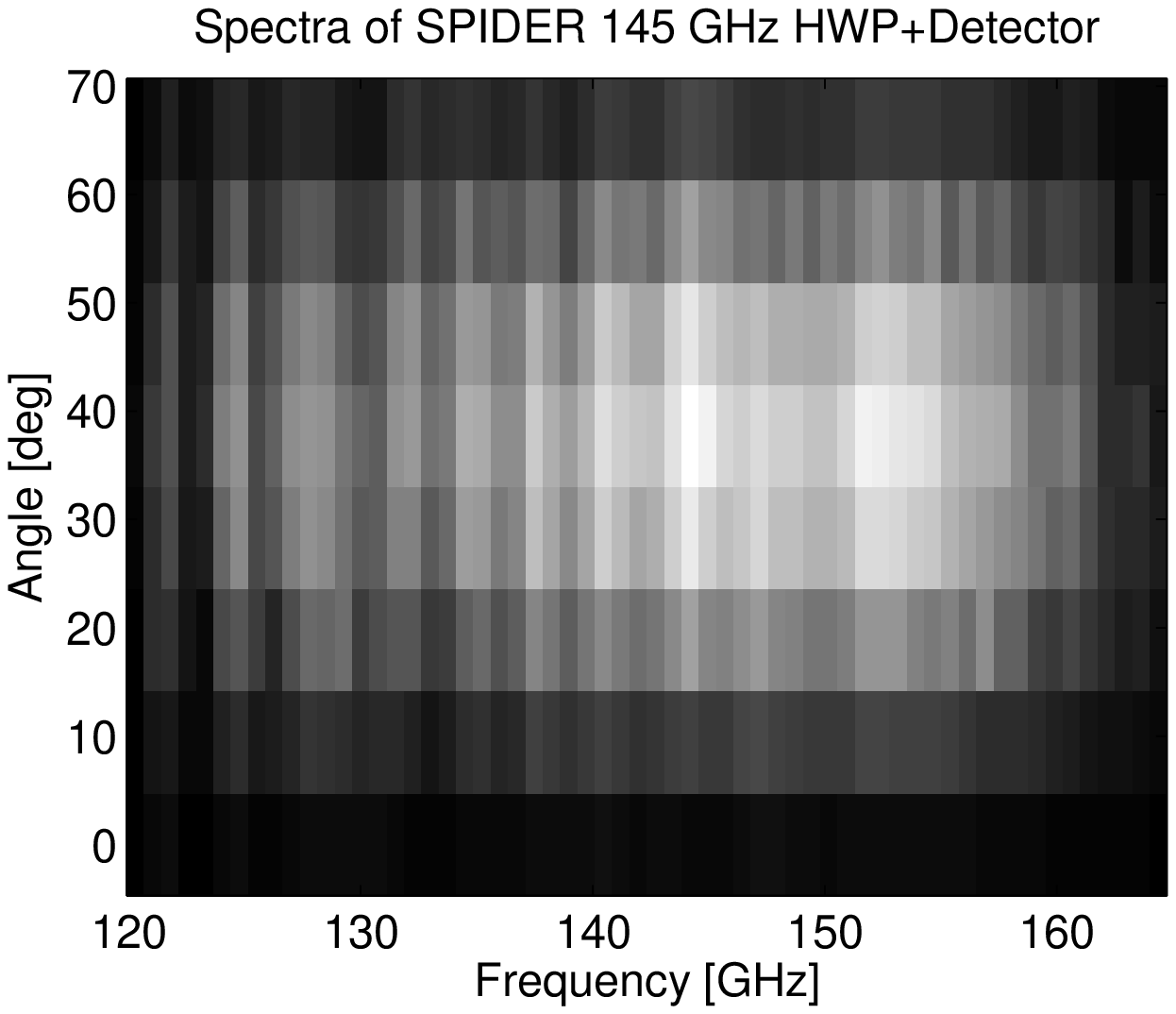}} 
\subfigure{\includegraphics[width=0.58\textwidth]{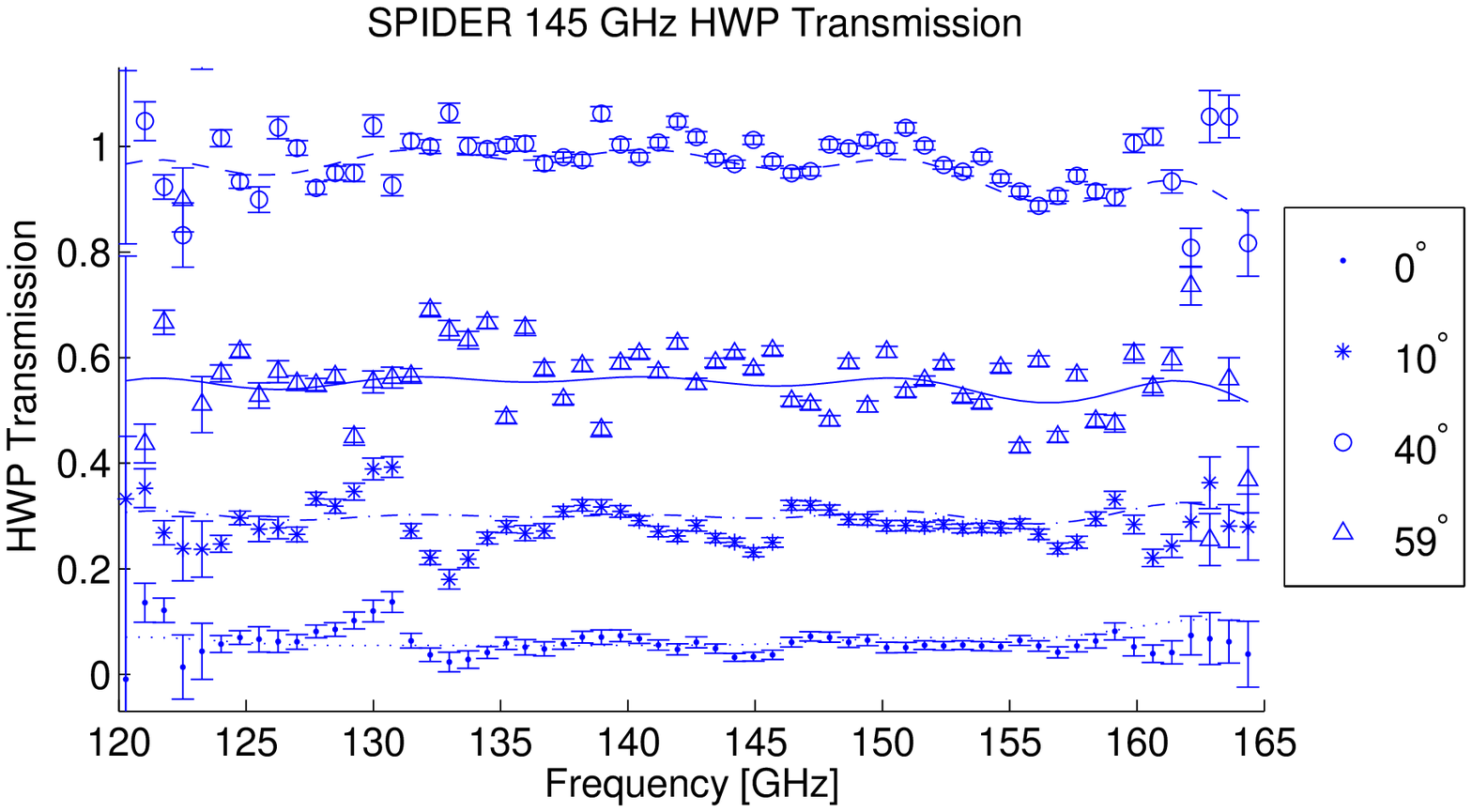}} 
\caption{Spectra taken with a single polarized detector of the cryogenic AR-coated HWP mounted in the prototype {\sc Spider} receiver. The left panel shows the spectra of the combined detector and HWP taken at eight HWP angles between 0 and $66^\circ$. The right panel shows a HWP transmission model (smooth curves) and data scaled by an estimated detector-only spectrum. Since we did cool down the instrument a second time to measure the detector-only spectrum, we combined the HWP+detector spectra to estimate the detector-only spectrum. \label{test_cryostat_spectra}}
\end{center}
\end{figure*}

We also mounted the prototype AR-coated HWP and its cryogenic rotation mechanism in the prototype {\sc Spider} receiver\cite{runyan10}, and took polarized FTS spectra with the HWP rotated to eight angles. The spectra from one polarized detector are plotted in Figure~\ref{test_cryostat_spectra}. Since at present we only have data from the combined HWP and detector, separating out the spectral properties of each is a challenge, but can be attempted by looking for the spectrum of the response that is independent of HWP angle and assigning that to the detector. We start by dividing the entire dataset by a fiducial HWP model. If this model is correct, dividing through leaves only the intrinsic response of the detector in the dataset. We keep the indices the same, but use an MCMC to vary the crystal axis angle $\theta_{0}$ in the HWP model until the remaining spectrum for each angle in the dataset is similar. This lets us combine data from the eight angles to produce an estimate of the detector response. We then divide the raw spectra by this estimated detector response to produce the estimated HWP-only spectra plotted in Figure~\ref{test_cryostat_spectra}. The observed polarization modulation agrees broadly with the fiducial model, but the scatter is larger than the noise estimate. This may be due to an incorrect estimate of the detector-only spectrum. Running the instrument again without the HWP would directly give a detector-only spectrum, and would allow us to straightforwardly determine the in-band HWP properties. Still, our early results show promising polarization modulation with the HWP that is broadly consistent with our model and our previous broadband lab testing.

\section{Modeling the HWP Response to Source Spectra}

Now that we have indices of refraction of the HWP materials, and since spectra of the integrated instrument are broadly consistent with the physical optics model of the HWP, we can make a projection of how the instrument will perform when observing the CMB and foregrounds on the sky. To do this, we have calculated the Mueller matrix of a {\sc Spider} 145 GHz HWP including the effects of the AR coat and reflections using the method discussed in Bryan, Montroy, and Ruhl \cite{bryan10}. A Mueller matrix is a 4-by-4 matrix of real numbers that maps the Stokes parameters of an incident polarization state onto the polarization state of the output light. In general, the band-averaged Mueller matrix of a single-plate HWP will have four independent non-zero elements.
\begin{equation}\label{m_hwp}
\mathsf{M}_{HWP} \equiv \left[ \begin{array}{cccc} T & \rho & 0 & 0 \\ \rho & T & 0 & 0  \\ 0 & 0 & c & -s \\ 0 & 0 & s & c \end{array} \right]
\end{equation}
Once calculated or estimated through instrument calibration, these four parameters completely characterize the polarized response of a single-plate HWP to a particular source spectrum. Since the CMB is not expected to be circularly-polarized, and the detectors have no sensitivity to circular polarization, the $s$ parameter is not relevant for CMB polarimetry. To perform the band-averaging calculation, we assumed a top-hat detector spectrum from 130 GHz to 160 GHz. As estimates for the source spectra, we took
\begin{align}
S(\nu) \propto 
\begin{cases} 
  1 & \mathrm{Flat} \\
  \frac{dB}{dT} (\nu,2.725~\mathrm{K})  & \mathrm{CMB} \\
  \nu^{1.67} B(\nu,9.6~\mathrm{K}) + 0.0935 \nu^{2.7} B(\nu,16.2~\mathrm{K}) & \mathrm{Dust} \\
  \nu^{-1} & \mathrm{Synchrotron} \\
  \nu^{-.14} & \mathrm{Free-free},
\end{cases}
\end{align}
where $B(\nu,T)$ is the blackbody function. The results are shown in Table \ref{hwp_params}.  The $c$ parameter deviates from ideality by almost $5\%$, which is a relatively large effect. This pushes us towards handling the non-idealities through calibration and a modified mapmaking algorithm, as described in Bryan, Montroy, and Ruhl \cite{bryan10}.

\begin{table}
\begin{center}
\caption{Calculated Mueller matrix elements for an optimized cryogenic sapphire HWP with a quartz AR-coat for the {\sc Spider} 145 GHz HWP. We optimized the HWP thickness based on our measured cold indices, and the AR-coat thickness based on the index in Lamb\cite{lamb95}. The first row shows the HWP parameters averaged within the {\sc Spider} 145 GHz passband. The CMB, Dust, Synchrotron, and Free-free rows all are band-averaged against the source spectra within the passband. The last row shows the parameter values of an ideal HWP for comparison. \label{hwp_params}}
\begin{tabular}{ c | c | c | c | c|}
 & $T$ & $\mathbf{\rho}$ & $c$  & $s$\\
\hline
\hline
\textbf{Flat} & 0.97389 & 0.01069 & -0.95578 & \phantom{-}0.00170 \\
\hline
\textbf{CMB} & 0.97396& 0.01069 & -0.95591 & -0.00952 \\
\hline
\textbf{Dust} & 0.97391 & 0.01080 & -0.95563 & -0.03598 \\
\hline
\textbf{Synchrotron} & 0.97382 & 0.01070 & -0.95565 & \phantom{-}0.01280\\
\hline
\textbf{Free-free} & 0.97388 & 0.01069 & -0.95577 & \phantom{-}0.00325 \\
\hline
\textrm{(Ideal HWP)} & 1 & 0 & -1 & 0\\
\hline
\end{tabular}
\end{center}
\end{table}

The variation of the parameters between the CMB and foreground sources is not as worrisome. For an optimized sapphire HWP with quartz AR coat, the parameters vary among the different sources by much less than 1\%. Determining if this will affect foreground subtraction for {\sc Spider} will require full simulations of the flight. As a rough estimate, {\sc Spider} is targeting $B$-modes at the $r\sim.03$ level, and we will remove dust foregrounds by combining maps from several frequencies. To reach our target $B$-mode sensitivity, our goal for dust foreground subtraction is to reduce it by $\sim 95\%$, so a 1\% HWP systematic in the final spectra would mean that foreground subtraction would still be dominated by incomplete dust removal. Next-generation experiments targeting $r<.01$ will require better foreground removal, and may be impacted by this effect.

\section{Conclusion}

Rotating the polarization sensitivity of the {\sc Spider} instrument using cryogenic HWPs will mitigate systematics and improve the fidelity of the CMB polarization maps. Characterization of the spectral properties of the sapphire for the HWP shows strong agreement between the model and observed spectra. Our measured index differences at room and cryogenic temperatures agree with previous measurements. Early results indicate that the HWP is performing well in the prototype {\sc Spider} receiver. We have calculated the Mueller matrix of the {\sc Spider} HWP for the CMB and foreground spectra. The results indicate that source-varying HWP systematics should not affect foreground removal from the CMB maps, but a full instrument simulation will be necessary to determine definitively if it will be a problem.

\section{Acknowledgements}
 
 {\sc Spider} is funded by NASA grant number NNX07AL64G, and is also supported by a generous gift from the Gordon and Betty Moore Foundation. WCJ acknowledges the support of the Alfred P. Sloan Foundation.

\bibliography{report}   
\bibliographystyle{spiebib}   

\end{document}